# Electromagnetic Duality Symmetry-Protected Dirac-Like Cones

Muxuan Yang[1], Dongyang Yan[1], Lei Gao[1,2,*], Wei Liu[3,*], Yun Lai[4], Yadong Xu[1],

Zhi Hong Hang[1], Jie Luo[1,*]

[1]School of Physical Science and Technology & Collaborative Innovation Center of Suzhou Nano Science and Technology & Jiangsu Key Laboratory of Frontier Material Physics and Devices, Soochow University, Suzhou 215006, China

[2]School of Optical and Electronic Information, Suzhou City University & Jiangsu (Suzhou) Key Laboratory of Biophotonics, Suzhou 215104, China

[3]College for Advanced Interdisciplinary Studies, National University of Defense Technology, Changsha, Hunan 410073, China

[4]National Laboratory of Solid State Microstructures, School of Physics, Collaborative Innovation Center of Advanced Microstructures and Jiangsu Physical Science Research Center, Nanjing University, Nanjing 210093, China

*Correspondence: leigao@suda.edu.cn (Lei Gao); wei.liu.pku@gmail.com (Wei Liu); luojie@suda.edu.cn (Jie Luo)

**Abstract**

Dirac-like cones, featuring conical linear dispersions intersecting with flat bands, typically arise from accidental degeneracy of multiple modes that requires precise tuning of material and structural parameters, inherently limiting their robustness and applications. In this work, by introducing electromagnetic duality symmetry into photonic crystals, we demonstrate the emergence of intrinsically robust deterministic Dirac-like cones. We show that such symmetry (achieved through either self-dual particles or non-self-dual particle clusters with duality-glide symmetry) enforces double degeneracies for band structures of photonic crystals. Furthermore, by harnessing the joint duality-structural symmetry, multiple deterministic Dirac-like cones exhibiting exceptional resilience to lattice size variations can be obtained. Our introduction of an extra symmetry into photonic crystals establishes a profound connection between duality symmetry and Dirac physics, providing a robust platform for advanced photonic band engineering.

## 1. Introduction

Dirac cones, a hallmark of graphene's electronic band structure, have revolutionized the understanding of electron transport, giving rise to phenomena ranging from quantum Hall effect to Klein tunneling [1, 2]. Intriguingly, analogous Dirac physics manifests in classical wave systems such as anisotropic structures [3] and photonic crystals (PhCs) [4, 5]. In particular, two-dimensional PhCs with graphene-like honeycomb lattices host symmetry-protected Dirac cones at Brillouin zone corners ($K$ and $K'$ points) [6-8]. Recent breakthroughs further reveal engineered Dirac-like cones at the Brillouin zone center ($\Gamma$ point), which are characterized by conical linear dispersions intersecting with flat bands (Fig. 1) [9-29]. These unique dispersions open new avenues for developing ultralow-loss zero-index materials, which exhibit extraordinary properties and enable a wide range of applications, including directive emission [9], cloaking and electromagnetic impurity-immunity [9, 22], photonic doping [21-23, 30], phase-free waveguide devices [24-28]. However, unlike structural symmetry-protected Dirac cones, such Dirac-like cones emerge from fragile accidental degeneracies [9-29] that rely on meticulous tuning of both material and structural parameters with extreme precisions. To be specific, a tiny perturbation in material or structural parameters would induce the gap opening [Fig. 1(a)] and thus break the Dirac-like cones, and such fragility inherently limits their robustness and applications.

In this work, we obtain intrinsically robust deterministic Dirac-like cones in three-dimensional PhCs, protected by electromagnetic duality symmetry—a fundamental invariance of Maxwell's equations under electric-magnetic field interchange [31-46]. While this symmetry underpins foundational concepts from quantum gauge theories to optical scattering phenomena, including helicity-conserving scattering [33], Kerker effect and directional scattering [34-38], and scattering invariance [39-41], its role in photonic band engineering has remained largely unexplored. We demonstrate that electromagnetic duality symmetry enforces rigorous double degeneracies in PhCs composed of self-dual particles or non-self-dual particle clusters with duality-glide symmetry. By further enhancing structural symmetry via cubic lattices, we obtain multiple deterministic Dirac-like cones that exhibit intrinsic stability against perturbations [Fig. 1(b)]. Our discovery has revealed a hidden dimension of freedom for tailoring photonic band structures, offering new flexibilities and opportunities for advanced electromagnetic wave manipulation.

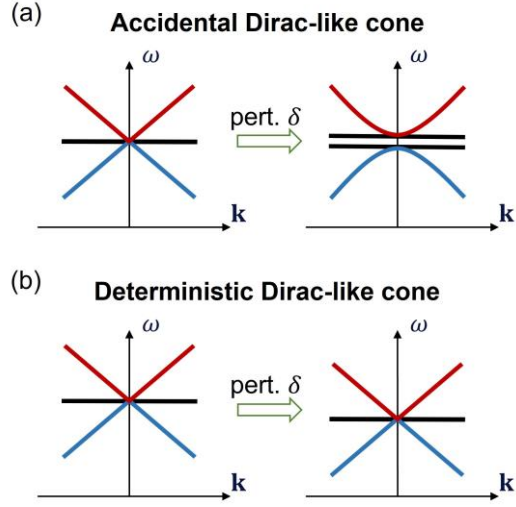

**Figure 1.** (a) Conventional Dirac-like cone arising from accidental degeneracy (left); Gap opening induced by a perturbation (pert.) $\delta$ in material or structural parameters (right). (b) Deterministic Dirac-like cone from symmetry-protected degeneracy (left); Intrinsic stability of the deterministic Dirac-like cone against perturbations while preserving the system symmetry, thus maintaining gapless states (right).

## 2. Electromagnetic Duality Transformation

We initiate our study by applying electromagnetic duality transformation to the fundamental constituent of a three-dimensional PhC, namely, particles characterized by relative permittivity $\varepsilon_r$ and relative permeability $\mu_r$. The electromagnetic duality transformation interchanges the roles of electric and magnetic fields via the duality transformation matrix $T(\alpha)$ through

$$\begin{pmatrix} \mathbf{E} \\ Z_0\mathbf{H} \end{pmatrix} \rightarrow \begin{pmatrix} \mathbf{E}' \\ Z_0\mathbf{H}' \end{pmatrix} = T(\alpha) \begin{pmatrix} \mathbf{E} \\ Z_0\mathbf{H} \end{pmatrix}, \tag{1}$$

$$\text{and } \begin{pmatrix} Z_0\mathbf{D} \\ \mathbf{B} \end{pmatrix} \rightarrow \begin{pmatrix} Z_0\mathbf{D}' \\ \mathbf{B}' \end{pmatrix} = T(\alpha) \begin{pmatrix} Z_0\mathbf{D} \\ \mathbf{B} \end{pmatrix}, \tag{2}$$

where $T(\alpha) = \begin{pmatrix} \cos\alpha & -\sin\alpha \\ \sin\alpha & \cos\alpha \end{pmatrix}$ with $\alpha$ being the transformation angle. The transformation matrix $T(\alpha)$ corresponds to a duality rotation with the rotation axis parallel to the propagation direction (along $\mathbf{E} \times \mathbf{H}$) [37, 44]. $Z_0$ is the characteristic impedance of free space. $\mathbf{E}$, $\mathbf{H}$, $\mathbf{D}$, and $\mathbf{B}$ represent the fields before transformation, while $\mathbf{E}'$, $\mathbf{H}'$, $\mathbf{D}'$, and $\mathbf{B}'$ denote the corresponding fields after transformation. The constitutive relation before transformation is given by

$$\begin{pmatrix} Z_0 \mathbf{D} \\ \mathbf{B} \end{pmatrix} = M \begin{pmatrix} \mathbf{E} \\ Z_0 \mathbf{H} \end{pmatrix} = c^{-1} \begin{pmatrix} \varepsilon_r & 0 \\ 0 & \mu_r \end{pmatrix} \begin{pmatrix} \mathbf{E} \\ Z_0 \mathbf{H} \end{pmatrix}, \tag{3}$$

where $c$ is the speed of light in free space. Utilizing Eqs. (1)-(3), the constitutive relation post-transformation is expressed as

$$\begin{pmatrix} Z_0 \mathbf{D}' \\ \mathbf{B}' \end{pmatrix} = T(\alpha) M T(\alpha)^{-1} \begin{pmatrix} \mathbf{E}' \\ Z_0 \mathbf{H}' \end{pmatrix} = M' \begin{pmatrix} \mathbf{E}' \\ Z_0 \mathbf{H}' \end{pmatrix}$$
$$= c^{-1} \begin{pmatrix} \varepsilon_r + (\mu_r - \varepsilon_r) \sin^2 \alpha & \frac{1}{2}(\varepsilon_r - \mu_r) \sin(2\alpha) \\ \frac{1}{2}(\varepsilon_r - \mu_r) \sin(2\alpha) & \varepsilon_r + (\mu_r - \varepsilon_r) \cos^2 \alpha \end{pmatrix} \begin{pmatrix} \mathbf{E}' \\ Z_0 \mathbf{H}' \end{pmatrix}. \tag{4}$$

The matrix $M'$ encapsulates the material parameters of the particle following the duality transformation. The diagonal elements of $M'$ correspond to the permittivity and permeability, while the off-diagonal elements describe magneto-electric coupling. Non-zero off-diagonal elements imply the necessity of bianisotropic materials, wherein electric and magnetic fields are intrinsically coupled [47]. However, the realization of such materials typically remains experimentally challenging [48].

Consequently, we restrict our analysis to scenarios where the off-diagonal terms vanish, leading to the condition $\frac{1}{2}(\varepsilon_r - \mu_r) \sin(2\alpha) = 0$. This condition yields two distinct solutions: (i) $\varepsilon_r = \mu_r$, and (ii) $\sin(2\alpha) = 0$. For the first solution ($\varepsilon_r = \mu_r$), the constitutive relation in Eq. (4) simplifies to

$$\begin{pmatrix} Z_0 \mathbf{D}' \\ \mathbf{B}' \end{pmatrix} = c^{-1} \begin{pmatrix} \varepsilon_r & 0 \\ 0 & \mu_r \end{pmatrix} \begin{pmatrix} \mathbf{E}' \\ Z_0 \mathbf{H}' \end{pmatrix}, \tag{5}$$

indicating that the constitutive relation remains invariant under the duality transformation, independent of transformation angle $\alpha$. Particles satisfying this condition are termed self-dual particles.

For the second solution ($\sin(2\alpha) = 0$), the constitutive relation in Eq. (4) reduces to

$$\begin{pmatrix} Z_0 \mathbf{D}' \\ \mathbf{B}' \end{pmatrix} = c^{-1} \begin{pmatrix} \mu_r & 0 \\ 0 & \varepsilon_r \end{pmatrix} \begin{pmatrix} \mathbf{E}' \\ Z_0 \mathbf{H}' \end{pmatrix}. \tag{6}$$

with $\alpha = \pm \pi/2$. This transformation corresponds to a parameter interchange $(\varepsilon_r, \mu_r) \to (\mu_r, \varepsilon_r)$ and simultaneous field interchanges

$$\begin{pmatrix} \mathbf{E} \\ Z_0 \mathbf{H} \end{pmatrix} \to \begin{pmatrix} \mathbf{E}' \\ Z_0 \mathbf{H}' \end{pmatrix} = \begin{pmatrix} -Z_0 \mathbf{H} \\ \mathbf{E} \end{pmatrix}, \tag{7}$$

$$\text{and } \begin{pmatrix} Z_0 \mathbf{D} \\ \mathbf{B} \end{pmatrix} \to \begin{pmatrix} Z_0 \mathbf{D}' \\ \mathbf{B}' \end{pmatrix} = \begin{pmatrix} -\mathbf{B} \\ Z_0 \mathbf{D} \end{pmatrix}. \tag{8}$$

Equations (6)-(8) indicates that one particle responding to electric (or magnetic) fields behaves identically to another particle with interchanged parameter $(\varepsilon_r, \mu_r) \to (\mu_r, \varepsilon_r)$ responding to magnetic (or electric) fields. Such particles are considered dual-paired and form a duality-symmetric particle cluster.

The above analysis shows that both duality-symmetric individual particles and particle clusters exhibit identical responses under the interchange of electric and magnetic fields. Notably, when these elements are organized into a periodic lattice to form a PhC, electromagnetic duality symmetry is anticipated to enforce mode degeneracies, manifested as the equivalence of two modes under electric-magnetic field interchange. This opens the possibility for realizing symmetry-protected Dirac-like cones. Based on this insight, we will demonstrate the emergence of electromagnetic duality symmetry-protected Dirac-like cones in two distinct classes of PhCs: one composed of self-dual particles and the other comprising non-self-dual particle clusters with duality-glide symmetry.

## 3. Self-Dual Particles for Deterministic Dirac-Like Cones

We first investigate a three-dimensional PhC composed of self-dual particles to elucidate the influence of electromagnetic duality symmetry on mode degeneracy. To isolate this effect from contributions due to structural symmetry, we utilize a cuboid lattice with distinct lattice constants $a_1$, $a_2$, and $a_3$ along the $x$, $y$, and $z$ directions, respectively. Figure 2(a) depicts the unit cell of this cuboid PhC, which contains a self-dual particle ($\varepsilon_r = \mu_r = 3$, radius $r_0$) at its center. The band structure of the PhC, with $a_1 = 4r_0$, $a_2 = 3.4r_0$, and $a_3 = 4.6r_0$, along the ΓX and ΓZ directions calculated using software COMSOL Multiphysics reveals a striking feature: all photonic bands possess doubly degenerate modes [Fig. 2(b)]—a phenomenon absent in conventional dielectric PhCs [9-29]. It is important to emphasize that this double degeneracy is not confined to the bands along the ΓX and ΓZ directions shown here. In fact, it pertains to all bands within the Brillouin zone. This finding strongly suggests that the degeneracy is a direct consequence of electromagnetic duality symmetry.

The origin of this double degeneracy can be understood through the geometric interpretation of duality transformation. For self-dual particles ($\varepsilon_r = \mu_r$), Eq. (4) tells that duality transformation of any $\alpha$ can be implemented. For our study, considering the orthogonality of degenerate modes within this cuboid-lattice PhC, we focus on $\alpha = \pm\pi/2$, corresponding to a parameter interchange $(\varepsilon_r, \mu_r) \to (\mu_r, \varepsilon_r)$ among particles and a simultaneous field interchange $(\mathbf{E}, Z_0\mathbf{H}) \to (-Z_0\mathbf{H}, \mathbf{E})$. Since all particles exhibit $\varepsilon_r = \mu_r$ (i.e., are self-dual), the parameter interchange does not alter the geometric configuration of the PhC. Nevertheless, the field interchange modifies the mode profiles, transforming one mode into another. The two modes respond identically within the same PhC, indicating that they are degenerate due to electromagnetic duality symmetry. This leads to a straightforward yet significant conclusion: an electric dipole (ED) mode with dipole moment $\mathbf{p}_i$ and a magnetic dipole (MD) mode with dipole moment $\mathbf{m}_i$ ($i = x, y, z$) are expected to exhibit degeneracy. This principle extends naturally to higher-order modes, underscoring the universality of this electromagnetic duality symmetry-protected double degeneracy.

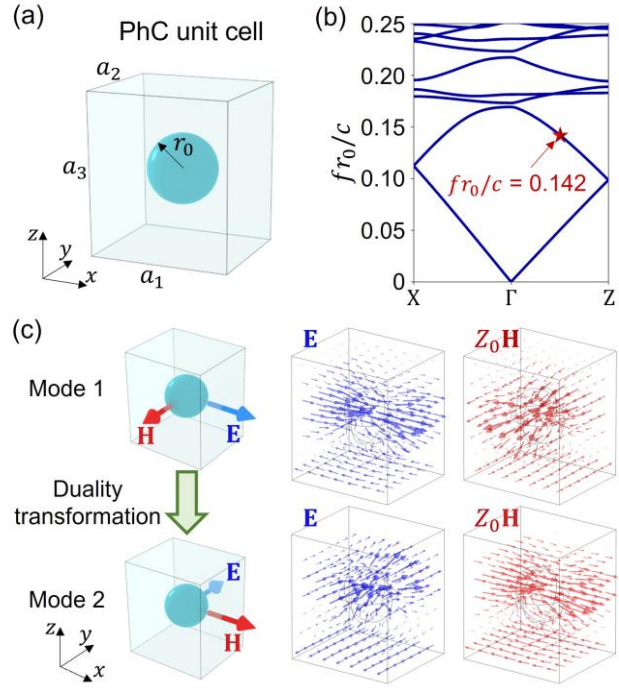

**Figure 2.** (a) Schematic of a PhC unit cell consisting of a self-dual particle ($\varepsilon_r = \mu_r = 3$, radius $r_0$) arranged in cuboid lattice with lattice constants $a_1$, $a_2$, and $a_3$ along the $x$, $y$, and $z$ directions, respectively. (b) Band structure for the cuboid-lattice PhC with $a_1 = 4r_0$, $a_2 = 3.4r_0$, and $a_3 = 4.6r_0$, showcasing doubly degenerate modes across all bands. (c) Left: illustration of two degenerate modes, i.e., mode 1 (upper) and mode 2 (lower), due to electromagnetic duality symmetry. Arrow maps of $\mathbf{E}$ (middle) and $Z_0\mathbf{H}$ (right) for the degenerate mode 1 (upper) and mode 2 (lower) at $\mathbf{k}_B = \hat{\mathbf{z}}0.5\pi/a_3$ and $fr_0/c = 0.142$, as marked by a star in (b).

Figure 2(c) exemplifies a general case of two degenerate modes: mode 1, characterized by an electric field **E** along the $x$ direction and a magnetic field **H** along the $-y$ direction, transforms into mode 2, with **E** along the $y$ direction and **H** along the $x$ direction, under the duality transformation. For numerical validation, we select the degenerate modes at the Bloch wave vector $\mathbf{k_B} = \hat{z} 0.5\pi/a_3$ and normalized frequency $fr_0/c = 0.142$, as indicated by a star in Fig. 2(b). The arrow maps of **E** (middle) and $Z_0\mathbf{H}$ (right) for the two degenerate modes, mode 1 (upper) and mode 2 (lower), are shown in Fig. 2(c). Clearly, we observe an interchange between the electric (or magnetic) field of one mode and the magnetic (or electric) field of its degenerate counterpart, in agreement with theoretical predictions. Note that the direction of either the electric or magnetic field is reversed, as implied in the duality transformation in Eqs. (6)-(8).

We emphasize that the duality transformation exchanges the electric and magnetic fields, rather than inducing a geometric rotation of these fields. The two degenerate modes do not display $\pi/2$ rotation symmetry on the $xy$ plane. Closer examination of the fields in Fig. 2(c) shows that the electric (or magnetic) field amplitude along the $x$ direction for one mode is greater than that along the $y$ direction for its degenerate counterpart. This behavior can be explained using perturbation theory [4]. According to this theory, a small perturbation $\Delta\varepsilon$ in permittivity induces a frequency shift $\Delta f$, as described by [4]

$$\Delta f = -\frac{f}{2} \frac{\int d^3\mathbf{r} \Delta\varepsilon(\mathbf{r})|\mathbf{E}(\mathbf{r})|^2}{\int d^3\mathbf{r} \varepsilon(\mathbf{r})|\mathbf{E}(\mathbf{r})|^2} + O(\Delta\varepsilon^2), \tag{9}$$

where $f$ and $\mathbf{E}(\mathbf{r})$ are the eigenfrequency and mode profile for the unperturbed case with a spatially distributed permittivity $\varepsilon(\mathbf{r})$. Given that the studied PhC has a larger lattice constant along the $x$ direction ($a_1$) compared to the $y$ direction ($a_2$), the electric field along the $x$ (or $y$) direction experiences a lower (or higher) effective permittivity due to the sparse (or dense) particle distribution. If the electric field were invariant under duality transformation, modes with electric field along the $x$ direction would have a higher eigenfrequency $f$. However, electromagnetic duality symmetry enforces identical eigenfrequencies, resulting in a greater electric field along the $x$ direction than along the $y$ direction. This perturbation theory also holds for magnetic fields. Consequently, we observe that the electromagnetic duality symmetry-protected degenerate modes have different field amplitudes, with the amplitude being smaller (or greater) along the direction with a smaller (or larger) lattice constant.

Next, we explore the interplay between electromagnetic duality symmetry and structural symmetry within a cubic lattice. This combination gives rise to fascinating mode degeneracies. Figure 3(a) illustrates an example, focusing on ED and MD modes. It is well established that three ED modes with dipole moments $\mathbf{p}_x$, $\mathbf{p}_y$, and $\mathbf{p}_z$ (or three MD modes with dipole moments $\mathbf{m}_x$, $\mathbf{m}_y$, and $\mathbf{m}_z$) are inherently degenerate at the $\Gamma$ point due to the $O_h$ symmetry of the cubic-lattice PhC [22, 23, 49]. Meanwhile, our earlier analysis has shown that electromagnetic duality symmetry enforces a degeneracy between an ED mode with dipole moment $\mathbf{p}_i$ and a MD mode with dipole moment $\mathbf{m}_i$ ($i = x, y, z$). Consequently, a deterministic Dirac-like cone emerges as a result of symmetry-protected sixfold degeneracy of ED and MD modes [Fig. 3(a)].

To numerically confirm the existence of this symmetry-protected Dirac-like cone, we study the mode evolution in a cubic-lattice PhC composed of self-dual particles with $\varepsilon_r = \mu_r = 3$ as a function of the lattice constant $a$, as shown in Fig. 3(b). The inset depicts the PhC unit cell. The black and red bands represent fourfold and sixfold degenerate modes, respectively. The first red band corresponds to the degenerate ED and MD modes, while the second red band stems from degenerate electric quadrupole (EQ) and magnetic quadrupole (MQ) modes, both contributing to the formation of Dirac-like cones. Figure 3(c) presents two cases: one PhC with $a = 2.5r_0$ (left) and another with $a = 4.5r_0$ (right). Both configurations exhibit multiple Dirac-like cones, identified by their characteristic conical linear dispersions intersecting with flat bands, as indicated by red circles. Notably, the conical linear dispersions persist along the $\Gamma$Y direction. In the first PhC, the Dirac-like cones observed at $fr_0/c = 0.203$ and $0.244$ result from degenerate dipole and quadrupole modes, respectively. These cones persist in the second PhC but are shifted to $fr_0/c = 0.171$ and $0.214$, respectively. Additionally, an extra Dirac-like cone emerges at a higher frequency $fr_0/c = 0.238$, associated with higher-order modes. These results demonstrate the robust nature of multiple Dirac-like cones against lattice size variations—a hallmark of symmetry protection [15], underscoring the emergence of multiple symmetry-protected Dirac-like cones.

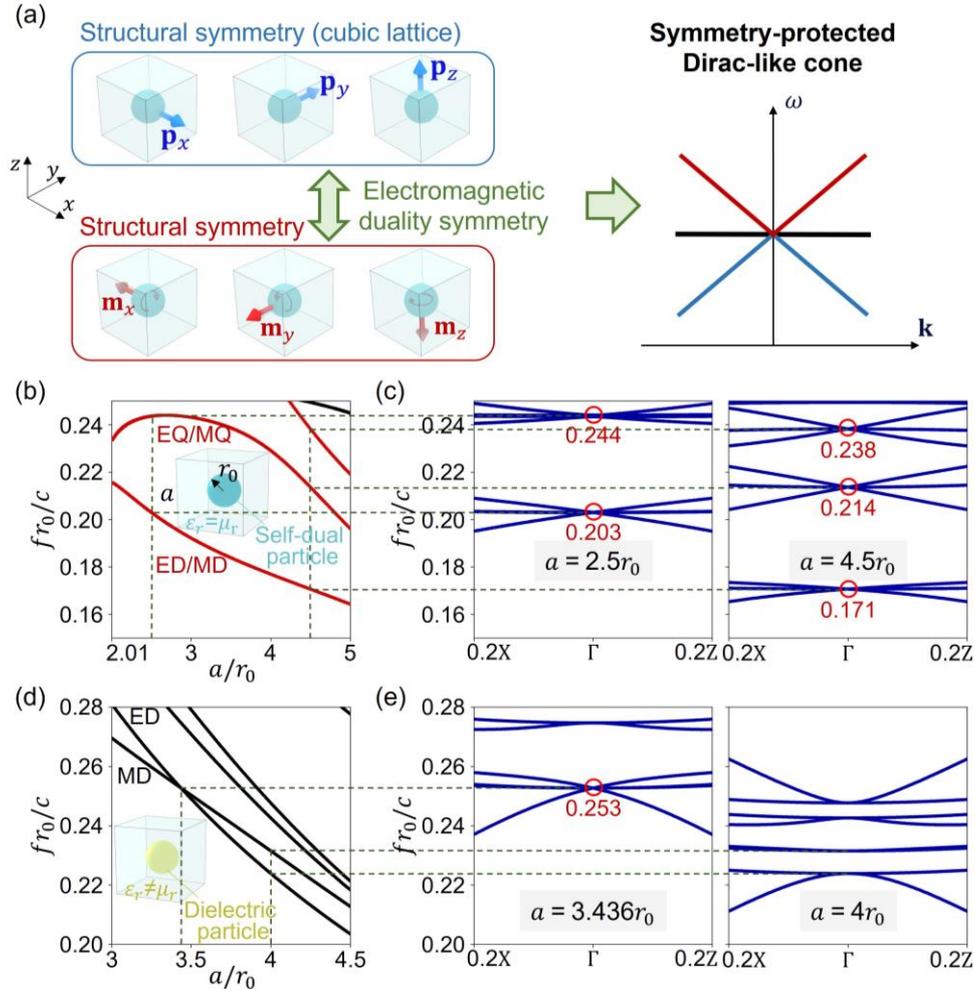

**Figure 3.** (a) A cubic-lattice PhC, composed of self-dual particles, exhibits a deterministic Dirac-like cone formed by symmetry-protected sixfold degeneracy: three ED modes (dipole moments $\mathbf{p}_x$, $\mathbf{p}_y$, and $\mathbf{p}_z$) and three MD modes (dipole moments $\mathbf{m}_x$, $\mathbf{m}_y$, and $\mathbf{m}_z$). The degeneracy of the three ED (or MD) modes arises from structural symmetry, while the degeneracy between an ED mode with dipole moment $\mathbf{p}_i$ and a MD mode with dipole moment $\mathbf{m}_i$ ($i = x, y, z$) is ensured by electromagnetic duality symmetry. (b) Mode evolution in the cubic-lattice duality-symmetric PhC as a function of $a$, with the black and red bands comprising fourfold and sixfold degenerate modes, respectively. The inset depicts the cubic PhC unit cell containing a self-dual particle with $\varepsilon_r = \mu_r = 3$. (c) Band structures for duality-symmetric PhCs with $a = 2.5r_0$ (left) and $a = 4.5r_0$ (right). (d) Mode evolution in a cubic-lattice dielectric PhC as a function of $a$, with the inset illustrating the unit cell containing a dielectric particle ($\varepsilon_r = 3$ and $\mu_r = 1$). (e) Band structures for dielectric PhCs with $a = 3.436r_0$ (left) and $a = 4r_0$ (right). The red circles in (c) and (e) indicate Dirac-like cones arising from symmetry-protected and accidental degeneracies, respectively.

For comparison, we examine a dielectric PhC composed of particles with $\varepsilon_r = 3$ and $\mu_r = 1$. In this case, the ED and MD modes are only coincidentally degenerate at a specific lattice size, i.e., $a = 3.436 r_0$ [Fig. 3(d)]. As a consequence, the resulting accidental Dirac-like cone is fragile. It appears only when $a = 3.436 r_0$ [Fig. 3(e), left] and vanishes when $a$ is altered, e.g., to $a = 4 r_0$ [Fig. 3(e), right]. Moreover, such accidental Dirac-like cones typically occur singly within a single PhC, and achieving multiple cones simultaneously is highly challenging [20]. This limitation stems from the fact that the formation of accidental Dirac-like cones requires tuning of material and structural parameters with extreme precisions, imposing substantial constraints. Notably, by harnessing the combined power of electromagnetic duality symmetry and structural symmetry, we not only achieve inherently robust symmetry-protected Dirac-like cones but also enable the realization of multiple such cones. In principle, when higher-order modes are taken into account, an unlimited number of Dirac-like cones can be created within a single duality-symmetric PhC.

## 4. Non-Self-Dual Particles with Duality-Structural Symmetry for Deterministic Dirac-Like Cones

In addition to self-dual particles characterized by $\varepsilon_r = \mu_r$, non-self-dual particle clusters that satisfy the parameter interchange $(\varepsilon_r, \mu_r) \to (\mu_r, \varepsilon_r)$ can also realize deterministic Dirac-like cones by exploiting duality and structural symmetries, as demonstrated below. To elucidate the role of electromagnetic duality symmetry in mode degeneracies, we first investigate a cuboid-lattice PhC. The unit cell of this PhC, illustrated in Fig. 4(a), consists of a non-self-dual particle cluster arranged in a cuboid lattice with lattice constants $2a_1$, $2a_2$, and $2a_3$ along the $x$, $y$, and $z$ directions, respectively. Each cluster consists of eight particles: four particles A ($\varepsilon_A = 9$, $\mu_A = 1$) and four particles B ($\varepsilon_B = 1$, $\mu_B = 9$), each with radius $r_0$, such that the parameter interchange $(\varepsilon_A, \mu_A) \to (\mu_B, \varepsilon_B)$ holds. These particles are arranged in a staggered configuration within a smaller cuboid lattice that is half the size of the PhC unit cell.

Similar to PhCs composed of self-dual particles, this system exhibits electromagnetic duality symmetry-protected double degeneracies. Figure 4(b) provides a geometric interpretation of this phenomenon. We begin by considering mode 1, where the electric and magnetic fields are reversed on the planes containing the lower and upper four particles (left panel). Applying a duality transformation $T(\pi/2)$ swaps particles A and B, accompanied by the field interchange $(\mathbf{E}, Z_0\mathbf{H}) \to (-Z_0\mathbf{H}, \mathbf{E})$ (middle panel). Due to the periodicity of the PhC unit cells, a glide transformation $G_z$: $(x, y, z) \to (x, y, z + a_3/2)$, corresponding to a half-period translation along $z$, restores the unit cell to its original configuration (right panel). However, the fields do not return to their initial state (mode 1). Instead, a notable field interchange between $\mathbf{E}$ and $\mathbf{H}$ occurs. This indicates that mode 1 is transformed into its degenerate counterpart, mode 2. Specifically, in the upper (or lower) half region of mode 1, the fields $\mathbf{E}$ and $\mathbf{H}$ exchange with fields $\mathbf{H}$ and $\mathbf{E}$ in the lower (or upper) half region of mode 2, with one of the components being reversed. Thus, by exploiting electromagnetic duality symmetry and glide symmetry, deterministic double degeneracies are achieved in such PhCs.

To numerically verify these findings, we calculate the band structure of the PhC with $a_1 = 4r_0$, $a_2 = 3.4r_0$, and $a_3 = 4.6r_0$ along the ΓX and ΓZ directions, as presented in Fig. 4(c). The results reveal that all photonic bands feature doubly degenerate modes. This double degeneracy pertains to all bands within the Brillouin zone. The middle panel provides an enlarged view for a detailed inspection of the double degeneracy. As an example, we select two degenerate modes at the Bloch wave vector $\mathbf{k_B} = \hat{z}0.001\pi/a_3$ and normalized frequency $fr_0/c = 0.099$, as marked by a star. The right panel displays the arrow maps of $\mathbf{E}$ (left) and $Z_0\mathbf{H}$ (right) for these degenerate modes. We observe that the fields within the upper (or lower) half region of mode 1 and those within the lower (or upper) half region of mode 2 satisfy the interchange rule $(\mathbf{E}, Z_0\mathbf{H}) \to (-Z_0\mathbf{H}, \mathbf{E})$, consistent with theoretical predictions.

We note that the two degenerate modes do not display $\pi/2$ rotation symmetry on the $xy$ plane. Similar to the mode characteristics in PhCs composed of self-dual particles (Fig. 2), the amplitude of fields along the direction with a smaller (or larger) lattice constant is smaller (or greater). Furthermore, the glide transformation here induces a spatial field shift in the $z$ direction. Consequently, the fields in the upper (or lower) half region of one mode correspond to those in the lower (or upper) half region of the other mode. Importantly, for modes far away from the Γ point in the ΓZ direction, additional phase changes must be taken into account for the degenerate modes.

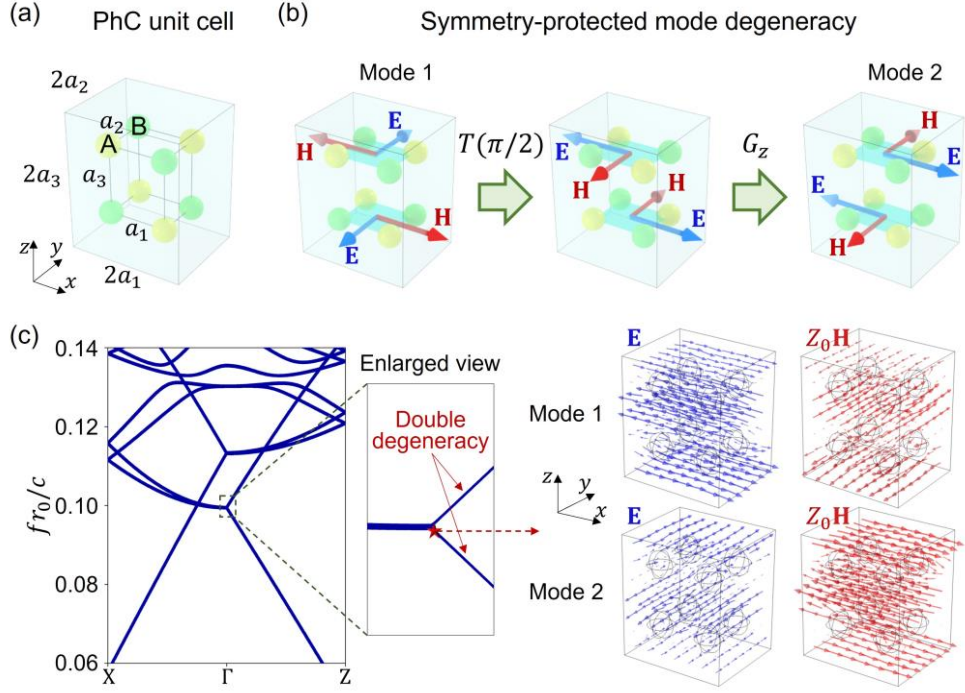

**Figure 4.** (a) Schematic of a PhC unit cell containing a non-self-dual particle cluster arranged in a cuboid lattice with lattice constants $2a_1$, $2a_2$, and $2a_3$ along the $x$, $y$, and $z$ directions, respectively. The particle cluster consists of eight particles: four particles A ($\varepsilon_A = 9$, $\mu_A = 1$) and four particles B ($\varepsilon_B = 1$, $\mu_B = 9$), each with radius $r_0$. These particles are arranged in a staggered configuration within a smaller cuboid lattice of half the size of the PhC unit cell. (b) Evolution of electric field **E** and magnetic field **H** under a duality transformation $T(\pi/2)$ and a glide transformation $G_z$, illustrating the symmetry-protected degeneracy of modes 1 and 2. (c) Band structure for the PhC with $a_1 = 4r_0$, $a_2 = 3.4r_0$, and $a_3 = 4.6r_0$ (left), with an enlarged view (middle). All bands comprise doubly degenerate modes. Right: arrow maps of **E** (left) and $Z_0\mathbf{H}$ (right) for two degenerate modes at $\mathbf{k_B} = \hat{\mathbf{z}}0.001\pi/a_3$ and $fr_0/c = 0.099$, as marked by a star in the lower band shown in the middle panel.

Our analysis of PhCs composed of self-dual particles (Fig. 3) suggests that intrinsically robust deterministic Dirac-like cones can be achieved by enhancing structural symmetry through cubic lattices. To demonstrate this, we consider a PhC unit cell containing a non-self-dual particle cluster arranged in a cubic lattice with lattice constant $2a$, as illustrated in Fig. 5(a). The particles are the same as those in Fig. 4 and maintain in a similar configuration but are positioned within a cubic lattice that is half the size of the PhC unit cell. Figure 5(b) shows the mode evolution in the PhC as a function of $a$. The black and red bands correspond to sixfold and twelvefold degenerate modes, respectively. The band structure for the PhC with $a = 4r_0$ is presented in Fig. 5(c), revealing two Dirac-like cones at $fr_0/c = 0.114$ and $0.151$, highlighted by red circles. These cones originate from the twelvefold degenerate modes. Moreover, we observe that these twelvefold degeneracies persist as the lattice size varies, indicating that these Dirac-like cones arise from the combined duality and spatial symmetries.

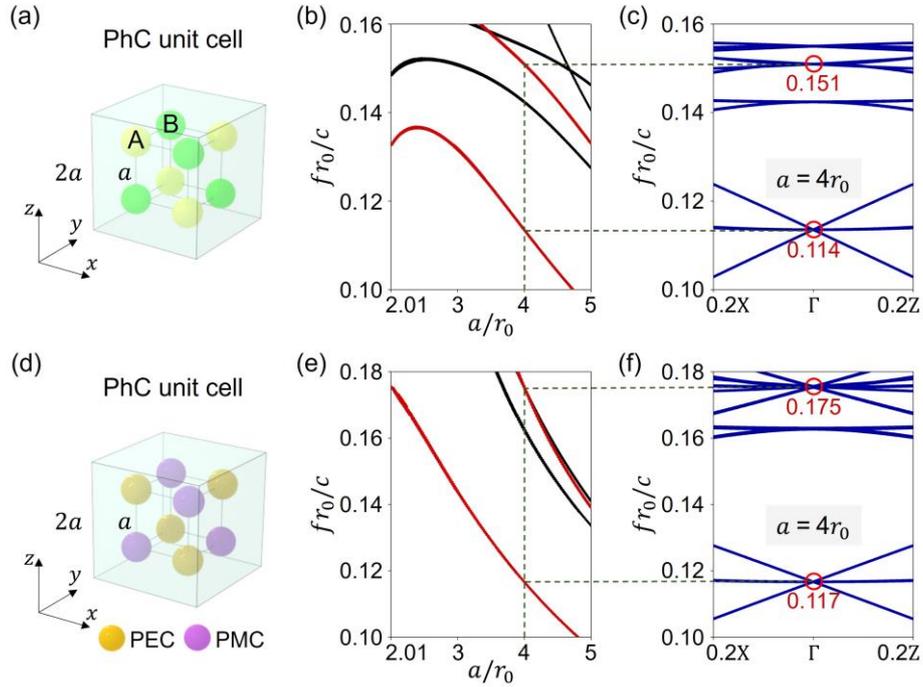

**Figure 5.** (a) Schematic of a PhC unit cell containing a non-self-dual particle cluster arranged in a cubic lattice with lattice constant $2a$. The particles are identical to those in Fig. 4 and maintain in a similar configuration but are positioned within a cubic lattice of half the size of the PhC unit cell. (b) Mode evolution in the PhC as a function of $a$. (c) Band structure for the PhC with $a = 4r_0$. (d) Schematic of a PhC unit cell containing PEC and PMC particles. The geometric configuration is identical to that in (a), with particles A and B replaced by PEC and PMC particles of identical dimensions, respectively. (e) Mode evolution in the PhC as a function of $a$. (f) Band structure for the PhC with $a = 4r_0$. In (b) and (e), the black and red bands correspond to sixfold and twelvefold degenerate modes, respectively. In (c) and (f), the red circles indicate the deterministic Dirac-like cones arising from symmetry-protected twelvefold degeneracies.
13

Such symmetry-protected Dirac-like cones can also be achieved using perfect electric conductors (PECs) and perfect magnetic conductors (PMCs), whose parameters satisfy the interchange $(\varepsilon_r, \mu_r) \to (\mu_r, \varepsilon_r)$ [44]. Figure 5(d) depicts a PhC unit cell composed of PEC and PMC particles. The geometric configuration is identical to that in Fig. 5(a), with particles A and B replaced by PEC and PMC particles of identical dimensions, respectively. The mode evolution as a function of $a$ confirms the inherent stability of the twelvefold degeneracies [Fig. 5(e), red bands], which correspond to symmetry-protected Dirac-like cones, as evidenced by the band structure for the PhC with $a = 4r_0$ [Fig. 5(f)]. These results showcase the existence of multiple deterministic Dirac-like cones as the consequence of the interplay between duality and structural symmetries.

## 5. Discussion and conclusion

In this study, we have introduced electromagnetic duality symmetry into PhCs for the first time, unlocking an extra degree of freedom for photonic band engineering. We have then demonstrated intrinsically robust deterministic Dirac-like cones in duality-symmetric PhCs, which are strictly protected by electromagnetic duality symmetry. Unlike conventional accidental Dirac-like cones that rely on meticulous tuning of material and structural parameters and typically occur singly, the symmetry-protected Dirac-like cones exhibit inherent robustness against lattice size variations. Moreover, as long as the system symmetry is preserved, an unlimited number of such cones can coexist within a single PhC.

The proposed framework of duality-symmetric PhCs establishes a versatile and robust platform for advanced photonic band engineering, with far-reaching implications for photonics and related disciplines. By strategically manipulating structural symmetry (e.g., through different lattice configurations), richer physical properties could be obtained. For instance, selectively reducing structural symmetry along specific directions could yield semi-Dirac cones [49, 50]. Although in this study, we focus on Dirac-like conical dispersions, the general band engineering principles unveiled are rooted in the interplay between duality and spatial symmetries, and thus hold transformative potentials for broad applications in the vibrant fields of topological, non-Hermitian and singular photonics [51-55].

## Acknowledgments

National Natural Science Foundation of China (Grant Nos. 12374293, 11974010, 12274313, 12274314, 12274315, 12474313, 12174188, 12474293); Natural Science Foundation of Jiangsu Province (Grant Nos. BK20221354, BK20221240, BK20233001); Suzhou Basic Research Project (Grant No. SJC2023003); Undergraduate Training Program for Innovation and Entrepreneurship, Soochow University (Grant No. 202310285027Z).

## Conflict of interest

The authors declare no conflict of interest.

## Data Availability Statement

The data that support the findings of this study are available from the corresponding author upon reasonable request.